# Fractal Growth on the Surface of a Planet and in Orbit around it

[1]Ioannis Haranas, [2]Ioannis Gkigkitzis, [3]Athanasios Alexiou

[1]*Dept. of Physics and Astronomy, York University,*
*4700 Keele Street, Toronto, Ontario, M3J 1P3, Canada*

[2]*Departments of Mathematics and Biomedical Physics,*
*East Carolina University, 124 Austin Building,*
*East Fifth Street, Greenville, NC 27858-4353, USA*

[3]*Department of Informatics, Ionian University,*
*Plateia Tsirigoti 7, Corfu, 49100, Greece*

**Abstract:**
Fractals are defined as geometric shapes that exhibit symmetry of scale. This simply implies that fractal is a shape that it would still look the same even if somebody could zoom in on one of its parts an infinite number of times. This property is also called self-similarity with several applications including nano-pharmacology and drug nanocarriers. We are interested in the study of the properties of fractal aggregates in a microgravity environment above an orbiting spacecraft. To model the effect we use a complete expression for the gravitational acceleration. In particular on the surface of the Earth the acceleration is corrected for the effect of oblateness and rotation. In the gravitational acceleration the effect of oblateness can be modeled with the inclusion of a term that contains the $J_2$ harmonic coefficient, as well as a term that depends on the square of angular velocity of the Earth. In orbit the acceleration of gravity at the point of the spacecraft is a function of the orbital elements and includes only in our case the $J_2$ harmonic since no Coriolis force is felt by the spacecraft. Using the fitting parameter $d = 3.0$ we have found that the aggregate monomer number $N$ is not significantly affected and exhibits a minute 0.0001% difference between the geocentric and areocentric latitudes of $90°$ and $0°$. Finally for circular and elliptical orbits around Earth and Mars of various inclinations and eccentricities the aggregate monomer number it's not affected at all at the orbital altitude of 300 km.

## Introduction:

A fractal can be defined as a geometric shape that has symmetry of scale. This simply implies that fractal is a shape that it would still look the same even if somebody could zoom in on one of its parts an infinite number of times. This property is also called self similarity. Fractals can be easily mathematically produced and can create detailed pictures of planets, waves, plants, and mountains. Fractal growth leads to aggregates with very unusual structures with potential applications in the field of solid state physics and materials (Slobodrian et al., 1991) and (Gauthier, et al., 1993). The aggregate mass $m_a$, consists of $N$ monomers of unit mass, that scale with the linear dimension $L$ in the following way:

$$M = N = \xi L^d \qquad (1)$$

where $\xi$ is a constant that depends on the topology of the aggregate, $d$ is the fractal dimension. The fractal dimension can be in the range $1 < d < 3$ but most of the time $2 < d < 3$ describes a slow aggregation process, which makes the density to scale according to the equation:

$$\rho \propto L^{d-3}. \tag{2}$$

Since $d < 3$ this implies that as $L$ increases the density $\rho$ decreases, which constitutes a very peculiar and unusual behaviour due to the fact that the density of matter remains constant. Furthermore, since the density $\rho$ decreases and the aggregate are growing implies the fractal aggregates become very tenuous when they grow large. In today's research fractal research extends even in the field of solar system. In a paper by Richardson (1995) the author gives a consistent numerical treatment for modelling fractal aggregate dynamics. Fractal aggregates play an important role in an number of astrophysical regimes, including the early solar system nebula and interstellar medium. Furthermore, in Krauss and Blum (2004) the authors use a microgravity experiment in order to form of dust agglomerates by Brownian motion induced collisions. They find that the agglomerates have fractal dimensions as low as 1.4. Similarly, in Dominik et al., (2006) the author discusses the results of laboratory measurements as well as theoretical models of dust aggregation in protoplanetary disks, as the initial step toward planet formation. In this paper we are interested to study the properties of fractal aggregates in a microgravity environment and in particular in an experiment that takes place in an experiment in orbit around the Earth or any other planetary body in the solar system.

## 2 A brief theory of fractals

Stoke's law predicts that a higher gravity results to smaller aggregates. Invoking the law of torque equality for aggregates of the same mass we can write:

$$\frac{L_1}{L_2} = \frac{g_2}{g_1}, \tag{3}$$

Where $L_1$ and $L_2$ are the corresponding linear dimensions and $g_1$ and $g_2$ are two different values of the acceleration of gravity could also suggest the behaviour that it's described by the equation below:

$$L(g) = \frac{c_0}{g} \tag{4}$$

where $c_0$ is a positive constant. For real aggregates with a given mass and discrete monomers we have that $r_{min} < L < r_{max}$, where a maximum radius $r_{max}$ implies a connected aggregates, a minimum radius $r_{min}$ of the length $L$ similarly indicates closed packed monomers. Following Deladurantaye et al., (1997) we modify Eq. (4) in the following way:

$$L(g) = r_{min} + \frac{\delta}{g(\phi, J_2, \omega_E) + \gamma} \tag{5}$$

where $g(\phi, J_2, \omega_E)$ is the acceleration of gravity as a function of geocentric latitude $\phi$, $J_2$ is the Earth oblateness coefficient also called $J_2$ harmonic coefficient, $\omega_E$ is the angular velocity of the Earth, and $\gamma$ and $\delta$ are fitting parameters. Substituting Eq. (5) into (1) we have that:

$$N = \xi \left[ r_{\min} + \frac{\delta}{(g(\phi, J_2, \omega_E) + \gamma)} \right]^d , \qquad (6)$$

Solving Eq. (6) for $d$

$$d(g) = \frac{\ln\left(\dfrac{N}{\xi}\right)}{\ln\left( r_{\min} + \dfrac{\delta}{(g(\phi, J_2, \omega_E) + \gamma)} \right)} , \qquad (7)$$

which is the fractal dimension of an aggregate with $N$ monomers, as a function of the acceleration of gravity that in general on the surface of the Earth is a function of the geocentric latitude $\phi$, oblateness coefficient also called $J_2$ harmonic coefficient, $\omega_E$ is the angular velocity of the Earth  We can use a closed packed of monomers to approximate the constants $\xi$ and $r_{min}$ and therefore $\xi = \dfrac{\pi}{3 r_{mon}^3 \sqrt{2}}$ , where $r_{mon}$ is the monomer radius, and $r_{\min} = \left( N / \xi \right)^{1/3} = \left( \dfrac{3\sqrt{2}}{\pi} N \right)^{1/3} r_{mon}$. Following Deladurantaye et al., (1997) and substituting for $\xi$ and $r_{min}$ into Eq. (7) and taking the limit as $g(\phi) \rightarrow \infty$, we obtain that $d(g) = 3$, which is the topological dimension. Similarly, $g(\phi) \rightarrow 0$, then $d$ tends asymptotically to a minimal value. Theory predicts that fractal aggregates increase with gravity. This is in agreement with the idea that gravity prevents dendrites from forming, allowing for a deeper penetration diffusion particles inside the aggregate increasing thus the fractal dimension.

## 3. Biological applications concerning drugs nanocarriers

Nowadays biological phenomena can be successfully modeled through the fractal dimensions and a large amount of biomedical applications, free of contaminants (Marin et al., 2013), have been already produced for the creation of skin, blood vessels, and nanoparticles for drug delivery or in vitro stem cells cultures for the treatment of neurodegenerative diseases. We can recall some basic mathematical notations concerning nanoparticles' properties such as surface, suspension and settling as follows (Sivasankar et al., 2010): For a spherical solid particle of diameter $d$, surface area per unit mass is given by the equation:

$$S_g = \left( \frac{\pi d^2}{4} \right) \left( \frac{\pi d^3 \rho_s}{6} \right)^{-1} = \frac{3}{2 d \rho_s} \qquad (8)$$

where $\rho_s$ is the solid density. Due to the small size of the nanoparticles, it is easy to keep them suspended in a liquid. Large micro-particles precipitate out more easily because of gravitational force, whereas the gravitational force is much smaller on a nanoparticle. Also, particle settling velocity is given by Stokes' law:

$$V = \frac{d^2 g \left( \rho_s - \rho_1 \right)}{18 \mu_1} \qquad (9)$$

where $g$ is gravitation acceleration (9.8 m/sec at sea level), $\rho_1$ is liquid density (997 kg/m3 for water at 25°C) and $\mu_1$ is viscosity (0.00089 Pa/sec for water at 25°C). In Haranas et al. (2012, 2013) the authors study the sedimentation velocity and concentration effect in a variable $g$ on the surface of a planetary body as well as in orbit above an orbiting spacecraft where corrections to the gravitational acceleration have been considered. Similarly, the authors give a Computational Study of the Mechanics of Gravity-induced Torque on Cells for a corrected Earth gravity as well as in orbit around earth. Additionally, Brownian fluctuations resist the particle settlement. According to Einstein's fluctuation-dissipation theory, average Brownian displacement $x$ in time $t$ is given as follow:

$$\chi = \sqrt{\frac{2 k_B T_t}{\pi \mu \, d}} \qquad (10)$$

where $k_B$ is the Boltzman constant (1.38x10$^{-23}$ J/K), and T is temperature in Kelvin.

The Brownian motion of a 1000nm particle due to thermal fluctuation in water is 1716 nm/sec, which is greater than the settling velocity of 430nm/sec. Hence, particles below 1000nm in size will not settle merely because of Brownian motion. This imparts an important property to nanoparticles that they can be easily kept suspended despite high solid density. For the nanoparticles, the gravitational pull is not stronger than the random thermal motion of the particles. Hence, nanoparticle suspensions do not settle which provides a long self-life.

While the morphology of drugs nanocarriers' surface seems to play a significant role in the functionality due to the occurred interactions and interfacial phenomena, the fractal analysis along with nonlinearity, scaling and chaos seems to illustrate in the more efficient way the production of biosimilar drugs with reduced side effects for the patients (Pippa et al., 2013 a, b, c, d, e). Under a variety of materials with different properties, drugs nanocarriers have been increased in the last decade due to their biocompatibility, biodegradability, flexibility, and minimal side effects (Marin et al., 2013). For instance nanoparticles made with biodegradable polymers have been already established as an important instrument in the treatment of neurodegenerative diseases, due to their ability to cross the blood-brain barrier and their high drug-loading capacity (Modi et al., 2010), or in the diagnosis and treatment of cardiovascular disease, due to their size, shape, and an available surface area for biomolecule conjugation (Godin et al., 2010). The main processes for nanoparticle purification include ultracentrifugation, dialysis, gel filtration, and cross-flow filtration (Pinto

Reis et al., 2006; Kowalczyk et al., 2011). Moreover, chemical factors such as pH can destabilize the nano-system while physical variables such as storage temperature, Brownian motion, and gravitational forces can produce aggregation, diffusion, or sedimentation of the colloidal particles (Abdelwahed et al., 2006). A suitable stabilizer or surfactant can be used in order to prevent aggregation or sedimentation phenomena (Biro et al., 2008).

There are cases where gravitational effects and interactions are used to produce effective organic pharmaceuticals with high impact in patient's treatment. We will mention in summary, two different methods with worth mentioned results. The first (Chen et al., 2004) uses a novel high-gravity reactive precipitation (HGRP) technique (Fig. 1) and the benzoic acid as a model compound in order to produce nanodrugs. Authors used the rotating packed bed method under high gravity, where rotating speed, reactant concentration and volume flow rate have been identified as key factors affecting the particle size.

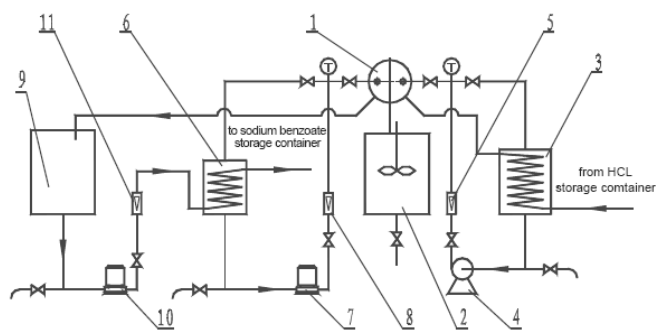

**Fig. 1** Schematic diagram of the high-gravity reactive precipitation set-up. (1) RPB, (2) benzoic acid storage container, sodium benzoate storage container, (4) sodium benzoate transportation pump, (5) sodium benzoate flowmeter, (6) HCl storage container, (7) HCl transportation pump, (8) HCl transportation flowmeter, (9) circulation water storage container, (10) circulation water transportation pump, (11) circulation water flowmeter. Reprinted from Feasibility of preparing nanodrugs by high-gravity reactive precipitation, 9/269, Chen JF, Zhou MY, Shao L, Wang YY, Yun J, Chew NY, Chan HK, *Int J Pharm*. 2004 Jan (1):267-74, with permission from Elsevier.

In the second case and according to a totally new publication study (Campbell et al., 2014), researchers in order to provide more efficient ways to reduce dosages and the frequency of injections, as well as to minimize the side effects of over-dosing, used neutron reflectometry in order to analyze the interaction of the liquid crystalline particles with a model cellular membrane whilst varying two parameters: gravity, in order to see how the interaction changed if the aggregates attacked the cell membrane from below as opposed to above and electrostatics in order to identify how the balance between the contrasting positive and negative charges of the aggregate and membrane affect the interaction (Fig. 2).

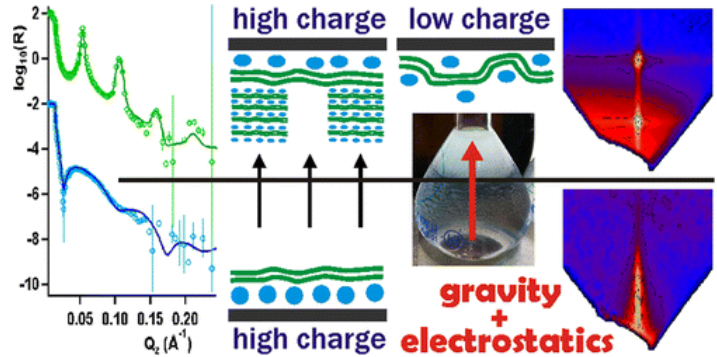

**Fig. 2** Gravity and Electrostatics are key factors regulating interactions between model cell membranes and self-assembled liquid crystalline aggregates of dendrimers and phospholipids. Reprinted from Key Factors Regulating the Mass Delivery of Macromolecules to Model Cell Membranes: Gravity and Electrostatics, Campbell RA, Watkins EB, Jagalski V, Åkesson-Runnsjö A, Cárdenas M, *ACS Macro Lett*. 2014, 3(2)121-125, with permission from the American Chemical Society.

## 4. The corrected gravitational acceleration on the surface of planetary body and in orbit around it

In our effort to calculate the effect of gravity in fractals let us consider the gravitational potential $V(r')$ at the surface of the Earth, where a fractal experiment is taking place under controlled conditions. We modify $g$ by writing it as the sum of three different potential terms namely:

$$V_{tot}(r') = -\frac{GM_E}{r'} + \frac{GM_E R_E^2 J_2}{2r'^3}\left(3\sin^2 \phi_E - 1\right) - \frac{1}{2}\omega_E^2 r'^2 \cos^2 \phi_E .$$ (11)

where the first term in the right-hand side (RHS) represents the central Newtonian potential, the second term represents the potential correction due to the $J_2 = C_{20}$ spherical harmonic of the Earth's gravitational potential (oblateness), and the third term represents the rotational potential, acting only on the surface of the Earth. Next, in Iorio (2011) the author analytically investigates the long term variations of the six orbital elements of a test particle in the presence of a non spherical rotating black hole with quadrupole moment $Q_2$ and angular momentum $S$ using a generic orientation of the spin axis $k$. Moreover, in Iorio (2012) the author numerically investigates the impact of the General Theory of Relativity (GTR) on the orbital part of the satellite-to-satellite range $\rho$ and range-rate $\dot{\rho}$ of the twin GRACE A/B spacecrafts through their post Newtonian (PN) dynamical equations of motion integrated in an Earth-centered frame over a time span $\Delta P = 1$d The author also computes the dynamical range and range-rate perturbations caused by the first six zonal harmonic coefficients $J_\ell$ $\ell = 2,3,4,5,6,7$ after expanding classically in multipole harmonics the Geopotential in order to evaluate their aliasing impact on the relativistic effects. Similarly, in Iorio et al. (2013) the authors use the asphericity of the Earth and employ higher zonal harmonic coefficients i.e. $\ell = 2,4,6$ and $m = 0$ to asses the error budget of the satellite LAGEOS and LAGEOS II by adopting an linear combination of the nodes $\Omega$ in both. Following Haranas, et al., (2012) we write the magnitude of the total of all accelerations on the surface of the Earth to be:

$$g_{tot} = -\frac{GM_E}{r'^2} + \frac{3GM_E R_E^2 J_2}{2r'^4}\left(3\sin^2\phi_E - 1\right) + \omega_E^2 r'\cos^2\phi_E \, , \qquad (12)$$

where $r'$ is the radial distance from the center of the Earth to an external surface point, $M_E$ is the mass of the Earth, $R_E$ is the radius of the Earth, $J_2$ is the zonal harmonic coefficient that describes the oblateness of the Earth, $\omega_E$ the angular velocity of the Earth, and $\phi_E$ is the geocentric latitude of the designed experiment. Zonal harmonics are simply bands of latitude, whose boundaries are the roots of a Legendre polynomial. This particular gravitational harmonic coefficient is a result of the Earth's shape and is about 1000 times larger than the next harmonic coefficient $J_3$ and its value is equal to $J_2 = $ -0.0010826269 (Kaula, 2000, Vallado, 2007). For the most recent Earth's global gravity field models the reader can consult the ICGEM or http://icgem.gfz-potsdam.de/ICGEM/. At the orbital point of the spacecraft the rotational potential on the surface of the earth does not affect the orbit of spacecraft. Therefore the gravitational acceleration as it is given by Equ. (2) can be transformed as a function of orbital elements, using standard transformations given by Kaula (2000) and Vallado (2007) namely $\sin\phi_E = \sin i \sin(u) = \cos\theta_E$, where, $\phi_E$ is the geocentric latitude, measured from the Earth's equator to the poles, and $\theta_E$ is the corresponding colatitude measured from the poles down to the equator ($\theta_E = 90 - \phi_E$), $u = \omega + f$ is the argument of latitude that defines the position of a body moving along a Kepler orbit, $i$ its orbital inclination, $\omega$ is the argument of the perigee of the spacecraft (not to be confused with angular velocity, which we write with subscripts – see nomenclature section below), $f$ is its true anomaly (an angle defined between the orbital position of the spacecraft and its perigee). The orbital elements are shown in figure 1 below:

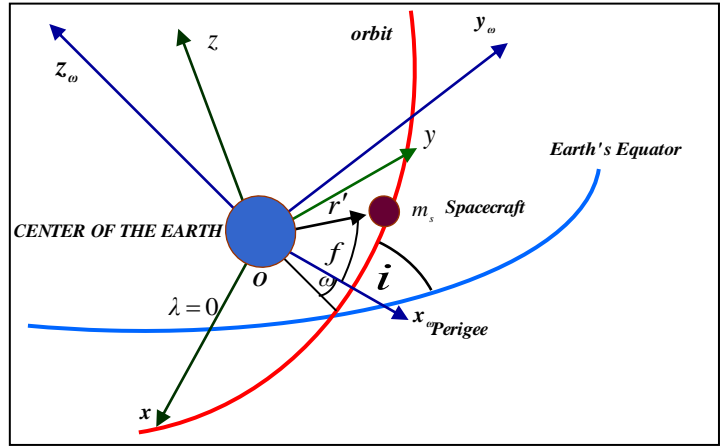

**Fig. 3** Explanation of the orbital elements: inclination $i$, argument of latitude $u = \omega + f$, and the radial vector $r'$ of the spacecraft, and $\lambda = 0$ is the zero longitude point on the Earth's equator, and $x_\omega, y_\omega, z_\omega$ define a right handed coordinate system. (Haranas et al. 2013)

To familiarize the reader with the orbital elements used here let us define the orbital elements appearing in our Eq. (4) below. $a_s$ is the semi-major axis that defines the size of the orbital ellipse. It is the distance from the center of the ellipse to an apsis i.e. the point where the radius vector is maximum or minimum (i.e. apogee and perigee points). Similarly, $e$ is the eccentricity, that defines the shape of the orbital ellipse (minor to major axis ratio), $i$ is the inclination of the orbit defined as the angle between the orbital and equatorial planes, and $\omega$ is the argument of the perigee, the angle between the direction of the ascending node (the point on the equatorial plane at which the satellite crosses from south to north) and the direction of the perigee. Finally, the true anomaly $f$, is the angle that locates the satellite in the orbital ellipse and is measured in the direction of motion from the perigee to the position vector of the satellite. Assuming an elliptical orbit the geocentric orbital distance $r'$ is given by (Vallado, 2007):

$$r' = \frac{a_s\left(1 - e^2\right)}{\left(1 + e\cos f\right)} \quad , \tag{13}$$

and therefore Eq. (2) becomes a function of the spacecraft orbital elements and therefore we have:

$$g_{tot} = \frac{GM_E\left(1 + e\cos f\right)^2}{a_s^2\left(1 - e^2\right)^2} - \frac{3GM_E R_E^2 J_2\left(1 + e\cos f\right)^4}{2a_s^4\left(1 - e^2\right)^4}\left(3\sin^2 i\sin^2 f - 1\right). \tag{14}$$

furthermore, we approximate the Earth as an ellipsoid of revolution so we can write that the radius of the Earth ellipsoid becomes (Kaula, 2000):

$$R_E = R_{eq}\left(1 - \left(f' + \frac{3}{2}f'^2\right)\sin^2\phi_E + \frac{3}{2}f'^2\sin^4\phi_E - ...\right) \approx R_{eq}\left(1 - f'\sin^2\phi_E\right) \tag{15}$$

where $R_{eq}$ is the equatorial radius of the Earth ellipsoid and where $f'$ is the Earth's flattening given by

$$f' = \frac{R_{eq} - R_{pol}}{R_{eq}} = \frac{3J_2}{2} + \frac{R_E^3 \omega_E^2}{2GM_E}$$ , (Stacey, 1977) where $R_{pol}$ is the Earth's polar radius.

## 4. Monomer number in planetary surface and space experiments

For experiments on the surface of the Earth the number of monomers can be written as a function of an acceleration of gravity $g$ that depends on the already mentioned parameters $\phi, J_2, \omega_E$, substituting Eq. (12) into (6) we obtain that $N$ is equal to:

$$N = \xi \left[ r_{min} + \frac{\delta}{\left( \gamma + \left( \frac{GM_E}{R_{eq}^2 \left(1 - f' \sin^2 \phi_E \right)} - \frac{3GM_E J_2}{2R_{eq}^2 \left(1 - f' \sin^2 \phi_E \right)} \left(3\sin^2 \phi_E - 1 \right) - \omega_E^2 R_{eq}^2 \left(1 - f' \sin^2 \phi_E \right) \cos^2 \phi_E \right) \right)} \right]^d , \quad (16)$$

Eq. (16) for the geocentric latitudes of $\phi_E = 0°, 45°, 90°$ becomes:

$$N = \xi \left[ r_{min} + \frac{\delta}{\left( \gamma + \left( \frac{GM_E}{R_{eq}^2} - \frac{3GM_E J_2}{2R_{eq}^2} - \omega_E^2 R_{eq}^2 \right) \right)} \right]^d , \quad (17)$$

$$N = \xi \left[ r_{min} + \frac{\delta}{\left( \gamma + \left( \frac{GM_E}{R_{eq}^2 \left(1 - \frac{f'}{2}\right)^2} - \frac{3GM_E J_2}{4R_{eq}^2 \left(1 - \frac{f'}{2}\right)^2} - \frac{1}{2}\left(1 - \frac{f'}{2}\right) R_{eq} \omega_E^2 \right) \right)} \right]^d , \quad (18)$$

$$N = \xi \left[ r_{min} + \frac{\delta}{\left( \gamma + \frac{GM_E}{R_{eq}^2 \left(1 - f'\right)^2} - \frac{3GM_E J_2}{R_{eq}^2 \left(1 - f'\right)^2} \right)} \right]^d . \quad (19)$$

In space, we first we consider circular orbits i.e. eccentricity $e = 0$ with inclinations $i = 0°$, 45° and 90°, and semimajor axis $a_s$. For circular orbits the true anomaly is undefined because the orbits do not possess a uniquely determined periapsis (perigee point), and therefore the *argument of latitude u* is used instead. In particular circular equatorial orbits $i = 0°$ and $e = 0$ do not have an ascending node point from which $u$ is defined from. In this case, Earth's equatorial plane is used as the reference plane, and the First Point of Aries as the origin of longitude and the *mean longitude* is used instead. This is the longitude that an orbiting body would have if its orbit were circular and its inclination $i = 0$. *Mean longitude* is given by the following relation $L = M + \Omega + \omega$, where $M$ is the mean anomaly defined as the angle between the perigee and the satellite radius vector assuming that the satellite moves with a constant angular velocity. Similarly, $\Omega$ is the right ascension of the node. On Earth it is measured positively (counter clockwise) in the equatorial plane from the longitude zero meridian ($\lambda = 0$) and the point of the orbit at which the satellite crosses the equator from south to north (ascending node). In this case Eq. (20) must have *mean longitude* used rather that argument of latitude. Therefore, substituting Eq. (11) in Eq. (6) we obtain:

$$N = \xi \left[ r_{min} + \frac{\delta}{\left( \gamma + \left( \frac{GM_E(1 + e\cos f)^2}{a_s^2(1 - e^2)^2} - \frac{3GM_E R_E^2 J_2 (1 + e\cos f)^4}{2a_s^4(1 - e^2)^4} \left( 3\sin^2 i \sin^2 f - 1 \right) \right) \right)} \right]^d , \quad (20)$$

taking $e = 0$ and $i = 0^\circ, 45^\circ, 90^\circ$ we obtain the following expressions:

$$N = \xi \left[ r_{min} + \frac{\delta}{\gamma + \frac{GM_E}{a_s^2} - \frac{3GM_E R_E^2 J_2}{2a_s^4}} \right]^d , \quad (21)$$

$$N = \xi \left[ r_{min} + \frac{\delta}{\gamma + \frac{GM_E}{a_s^2} - \frac{3GM_E R_E^2 J_2}{2a_s^4} \left( \frac{3\sin^2 u}{2} - 1 \right)} \right]^d , \quad (22)$$

$$N = \xi \left[ r_{min} + \frac{\delta}{\gamma + \frac{GM_E}{a_s^2} - \frac{3GM_E R_E^2 J_2}{2a_s^4} \left( 3\sin^2 u - 1 \right)} \right]^d . \quad (23)$$

Similarly, for elliptical orbits of the same inclination we obtain:

$$N = \xi \left[ r_{min} + \cfrac{\delta}{\left( \gamma + \left( \cfrac{GM_E(1+e\cos f)^2}{a_s^2(1-e^2)^2} - \cfrac{3GM_E R_E^2 J_2 (1+e\cos f)^4}{2a_s^4(1-e^2)^4} \left(3\sin^2 i \sin^2 f - 1\right) \right) \right)} \right]^d , \qquad (24)$$

$$N = \xi \left[ r_{min} + \cfrac{\delta}{\left( \gamma + \left( \cfrac{GM_E(1+e\cos f)^2}{a_s^2(1-e^2)^2} - \cfrac{3GM_E R_E^2 J_2 (1+e\cos f)^4}{2a_s^4(1-e^2)^4} \right) \right)} \right]^d , \qquad (25)$$

$$N = \xi \left[ r_{min} + \cfrac{\delta}{\left( \gamma + \left( \cfrac{GM_E(1+e\cos f)^2}{a_s^2(1-e^2)^2} - \cfrac{3GM_E R_E^2 J_2 (1+e\cos f)^4}{2a_s^4(1-e^2)^4} \left( \cfrac{3\sin^2 f}{2} - 1 \right) \right) \right)} \right]^d , \qquad (26)$$

$$N = \xi \left[ r_{min} + \cfrac{\delta}{\left( \gamma + \left( \cfrac{GM_E(1+e\cos f)^2}{a_s^2(1-e^2)^2} - \cfrac{3GM_E R_E^2 J_2 (1+e\cos f)^4}{2a_s^4(1-e^2)^4} \left(3\sin^2 f - 1\right) \right) \right)} \right]^d . \qquad (27)$$

## 5. Recovering planetary parameters from fractal planetary surface and space experiments

Today's methods for the modelling and calculation of various geophysical parameters involve the continuous acceleration component monitoring of orbiting satellites. In this section we propose the use of fractal experiments for the calculation at *least in principle* of parameters such as the $J_2$. As a first step, we want to establish analytical mathematical expressions that relate the number experimental parameters to the $J_2$ harmonic coefficient. A continuous monitoring of the fractal parameter $N$, $\xi$, and $r_{min}$ and the estimation of the constants $\gamma$, $\delta$ for a spacecraft fractal experiment can at least in principle result in the calculation of the spherical harmonic coefficient $J_2$. For that we use Eqs. (16) and (17) that is a modified equation for the number of monomer for an experiment that takes place aboard an orbiting spacecraft. Solving Eqs. (16) and (20) for $J_2$ and simplifying we obtain:

$$J_2 = \frac{1}{3GM_E\left((N/\xi)^{1/d} - r_{\min}\right)(3\cos 2\phi_E - 1)} + \frac{R_{eq}^2}{2}\left[\begin{array}{l} -4\left(GMr_{\min} + R_{eq}^2\left(\gamma\, r_{\min} + \delta\right) - 4\left(GM_E R_{eq}^2\gamma\right)(N/\xi)^{1/d}\right) \\[2mm] \left(\begin{array}{l} -8f'(\gamma r_{\min} + \delta)\sin^2\phi_E\left(f'\sin^2\phi_E - 2\right) \\ +8r_{\min}R_{eq}\omega_E^2\cos^2\phi_E\left(1 + f'\sin^4\phi_E\right)\left(3 - f'\sin^2\phi_E\right) \\ +(N/\xi)^{1/d}\left(\begin{array}{l} -R_{eq}\omega_E^2\cos^2\phi_E\left(2 - f'\left(1 - \cos 2\phi_E\right)^2\right) \\ +8f'\gamma\sin^2\phi_E\left(f'\sin^2\phi_E - 2\right) - 6f'r_{\min}R_{eq}^2\omega_E^2\sin^2 2\phi_E\right) \end{array}\right) \end{array}\right) \end{array}\right]$$

(28)

$$J_2 = \frac{a_s^2\left(1 - e^2\right)}{3GM_E R_E^2\left(\left(\dfrac{N}{\xi}\right)^{1/d} - r_{min}\right)\left(1 + e\cos f\right)^4\left(3\sin^2 f\sin^2 i - 1\right)}\left(\begin{array}{l} GM_E\left(2 + e^2\right)^2\left(\left(\dfrac{N}{\xi}\right)^{1/d} - r_{\min}\right) + 2a_s^2\left(1 - e^2\right)^2\left(\gamma\left(\dfrac{N}{\xi}\right)^{1/d} - \delta - \gamma\, r_{\min}\right) \\[4mm] + eGM_E\left(\left(\dfrac{N}{\xi}\right)^{1/d} - r_{\min}\right)\left(4\cos f + e\cos 2f\right) \end{array}\right)$$

(29)

## 6. Discussion and numerical results

In order to validate our results let us first calculate the values of the acceleration of gravity of the Earth's surface at the following geocentric latitudes namely $\phi_E = 0°, 30°, 45°, 60°, 90°$. For the Earth we use the following planetary parameters: $M_E = 5.9742\times10^{24}$ kg, $R_E = 63781363$ km, $G = 6.67\times10^{-11}$ m$^3$ kg$^{-1}$s$^{-2}$, $\omega_E = 7.292115486\times10^{-5}$ rad s$^{-1}$ and $J_{2_M} = 0.001082626$ (Vallado and McClain, 2007). Similarly for Mars we use: $M_M = 6.4191\times10^{23}$ kg, $R_M = 3397.2$ km, $G = 6.67\times10^{-11}$ m$^3$ kg$^{-1}$s$^{-2}$, $\omega_M = 7.0854\times10^{-5}$ rad s$^{-1}$ and $J_{2M} = 0.001964$ (Vallado and McClain, 2007). Furthermore, we need to calculate $\xi = \dfrac{\pi}{3r_{mon}^3\sqrt{2}}$ and $r_{\min} = (N/\xi)^{1/3} = (600/\xi)^{1/3}$. With reference to Peladurantaye et al. (1997) we use $0.5\ \mu\text{m} \le r_{mon} \le 5\ \mu\text{m}$, we obtain the following parameters which implies that $5.924\times10^{18}\text{m}^{-3} \le \xi \le 5.924\times10^{15}\text{m}^{-3}$. Similarly, we calculate that $r_{\min}$ is in the range $4.661\mu\text{m} \le r_{\min} \le 44.613\mu\text{m}$.

**Table 1** Parameters calculated for our numerical evaluation

| Sperical Monomer Radius $r_{mo}$ [μm] | $\xi$ [m$^{-3}$] | Minimum Monomer Radius $r_{min}$ [μm] |
|---|---|---|
| 0.5 | 5.924×10$^{18}$ | 4.661 |
| 5.0 | 5.924×10$^{15}$ | 46.613 |

For the calculation of the fitting constants $\gamma$ and $\delta$ we have used a number of monomers $N = 600$, and 1000 and $d = 3$. Therefore using Eq. (6) we solve the following system of equations:

$$(600)^{1/3} = \left(4.6613 \times 10^{-6} + \frac{\delta}{\gamma - 9.68136}\right) \qquad (30)$$

$$(1000)^{1/3} = \left(5.5266 \times 10^{-6} + \frac{\delta}{\gamma - 9.68136}\right) \qquad (31)$$

for $\gamma$ and $\delta$ we obtain that $\gamma \cong 9.68140$ and $\delta = -2.04753 \times 10^{-14}$ respectively the changes the number of monomers $N$ is tabulated in tables 1 to 6 below.

**Table 1** Number of monomers under constant and corrected acceleration of gravity as a function of geocentric latitude $\phi$ and of and fractal dimension of an aggregate with $N$ monomers $d$= 2.989 and 3.0.

| Number of monomers at constant acceleration of gravity with centripetal force, $J_2$ harmonic and latitude dependence excluded $N$ | Number of monomers $N$ for the corrected accelerations of gravity with centripetal force, $J_2$ harmonic and latitude dependence included at the given geocentric latitudes | |
| --- | --- | --- |
| | $\phi_{Earth}$ | $N$ |
| Fractal dimension $d$ = 3.0 | 0° | 599.979 |
| | 30° | 599.979 |
| 599.979 | 60° | 599.980 |
| | 90° | 599.980 |
| Fractal dimension $d$= 2.989 | 0° | 599.979 |
| | 30° | 599.979 |
| 686.725 | 60° | 599.980 |
| | 90° | 599.980 |

**Table 2** Number of monomers under constant and corrected acceleration of gravity as a function of areocentric latitude $\phi$ and of and fractal dimension of an aggregate with $N$ monomers $d$= 2.989 and 3.0

| Number of monomers $N$ at constant acceleration of gravity with centripetal force, $J_2$ harmonic and latitude dependence excluded $N$ | Number of monomers $N$ for the corrected accelerations of gravity with centripetal force, $J_2$ harmonic and latitude dependence included at the given geocentric latitudes | |
| --- | --- | --- |
| | $\phi_{Mars}$ | $N$ |
| Fractal dimension $d$ = 3.0 | 0° | 599.979 |
| | 30° | 599.979 |
| 599.979 | 60° | 599.980 |
| | 90° | 599.980 |
| Fractal dimension $d$ = 2.989 | 0° | 599.979 |
| | 30° | 599.979 |
| 686.725 | 60° | 599.980 |
| | 90° | 599.980 |

Next let us assume that a fractal structure experiment takes place above an orbiting spacecraft orbiting Earth and Mars respectively at an altitude $h$ = 300 km. First we consider circular orbits i.e. eccentricity $e$ = 0 and inclinations $i$ = 0, 45, 90 respectively, and then elliptical orbits of various eccentricities $e$ = 0, 0.01, 0.1, 0.7. Our results are tabulated in tables (3) and (4) below:

**Table 3** *N*umber of monomers *N* under constant and corrected orbital acceleration of gravity for a spacecraft orbiting Earth at $h$ = 300 km with eccentricity $e$ = 0, 0.01, 0.1, 0.7 various orbital inclinations $i$, and fitting parameters $d$ = 2.989 and 3.0 correspondingly.

| Number of monomers *N* at constant acceleration of gravity with centripetal force, $J_2$ harmonic and latitude dependence excluded *N* | Number of Monomers *N* for the Corrected accelerations of gravity with centripetal force, $J_2$ harmonic and latitude dependence included at the given spacecraft orbital inclinations and eccentricities $e$ = 0, 0.01, 0.1, 0.7 respectively. | |
|---|---|---|
| | *i* | *N* |
| Fractal dimension $d$ = 3.0 | 0° | 599.979 |
| | 30° | 599.979 |
| 599.979 | 60° | 599.979 |
| | 90° | 599.979 |
| Fractal dimension $d$ = 2.989 | 0° | 686.725 |
| | 30° | 686.725 |
| 686.725 | 60° | 686.725 |
| | 90° | 686.725 |

**Table 4** *N*umber of monomers *N* under constant and corrected orbital acceleration of gravity for a spacecraft orbiting Mars at $h$ = 300 km with eccentricity $e$ = 0, 0.01, 0.1, 0.7 various orbital inclinations $i$, and fitting parameters $\delta$ = 2.989 and 3.0 correspondingly.

| Number of monomers *N* at constant acceleration of gravity with centripetal force, $J_2$ harmonic and latitude dependence excluded *N* | Number of monomers *N* for the corrected accelerations of gravity with centripetal force, $J_2$ harmonic and latitude dependence included at the given spacecraft orbital inclinations and eccentricities $e$ = 0, 0.01, 0.1, 0.7 respectively. | |
|---|---|---|
| | *i* | *N* |
| Fractal dimension $d$ = 3.0 | 0° | 599.979 |
| | 30° | 599.979 |
| 599.979 | 60° | 599.979 |
| | 90° | 599.979 |
| Fractal dimension $d$ = 2.989 | 0° | 686.725 |
| | 30° | 686.725 |
| 686.725 | 60° | 686.725 |
| | 90° | 686.725 |

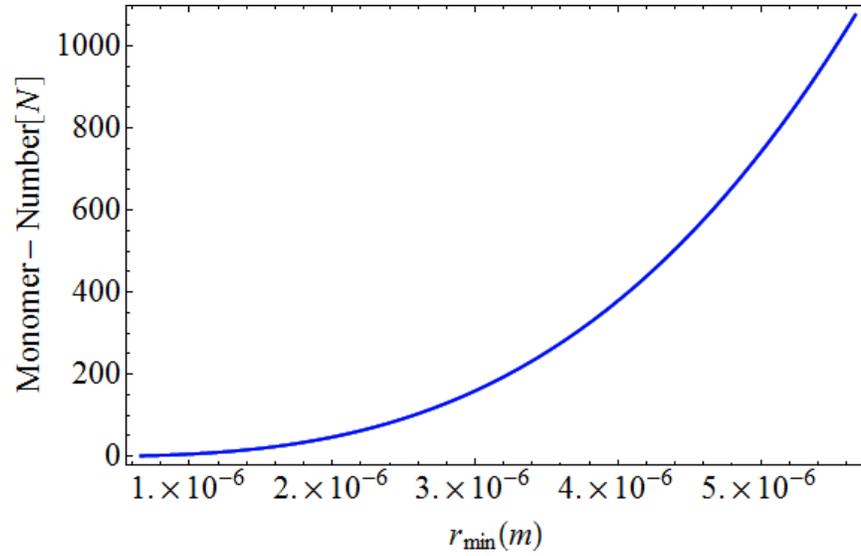

**Fig. 4** Monomer number as function closed packed monomers distance $r_{min}$ and a fractal dimension parameter $d = 3.0$ for an experiment taking place at geocentric latitude $\phi = 90^o$ on the surface of the Earth.

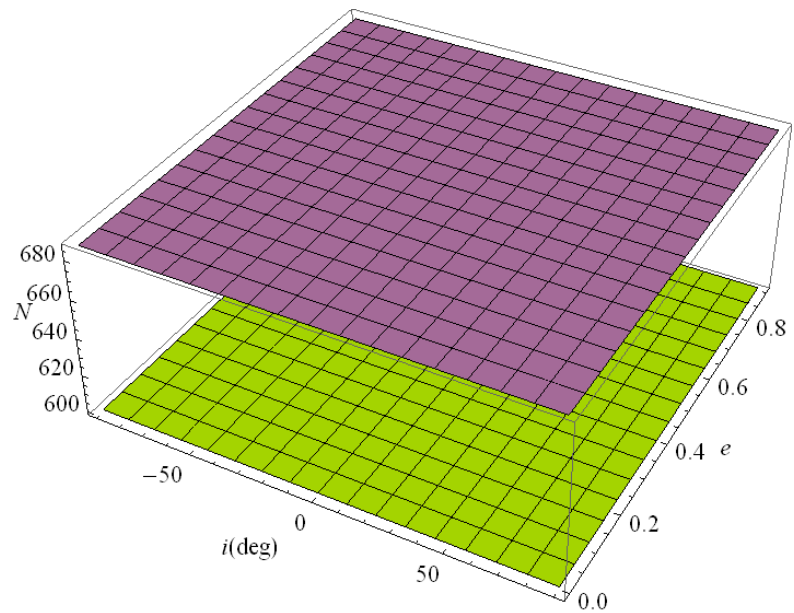

**Fig.5** Monomer number $N$ as a function of orbital inclination $i$ and eccentricity $e$ for two values of the fractal dimension parameter $d = 2.989$, 3.0 respectively. Green plane corresponds to $d = 3.0$ and purple plane corresponds to $d = 2.989$. This is for an experiment taking place 300 km above the surface of the Earth.

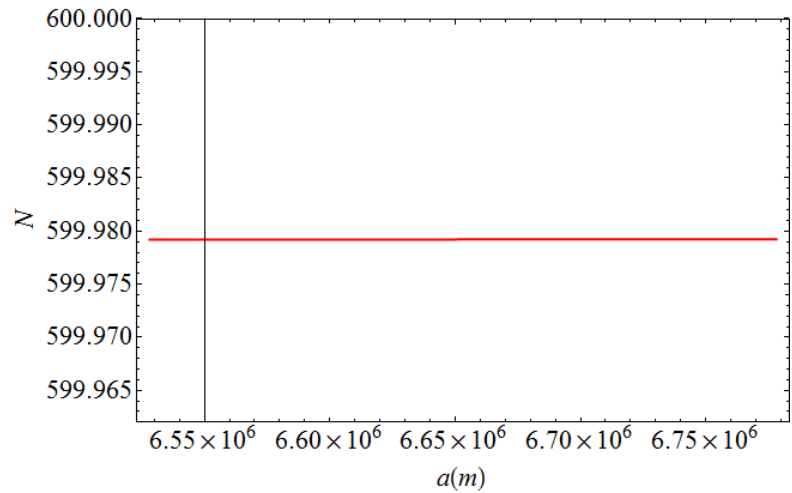

**Fig.6** Monomer number $N$ as a function of orbital semimajor axis $a$ for an orbit of eccentricity $e = 0.1$ and inclination $i = 90°$ for a fractal dimension parameter $d = 3.0$.

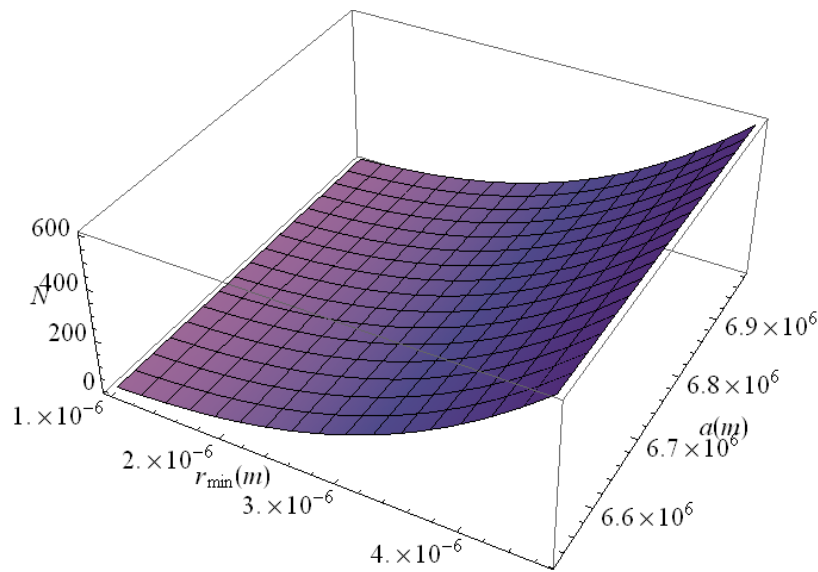

**Fig.7** Monomers number $N$ as a function of spacecraft orbital semimajor axis $a$ and monomer minimum distance $r_{min}$, for an experiment taking place in a spacecraft 300 km above the surface of the Earth in an elliptical polar orbit of eccentricity $e = 0.1$ and inclination $i = 90°$ and for a fractal dimension parameter $d = 3.0$.

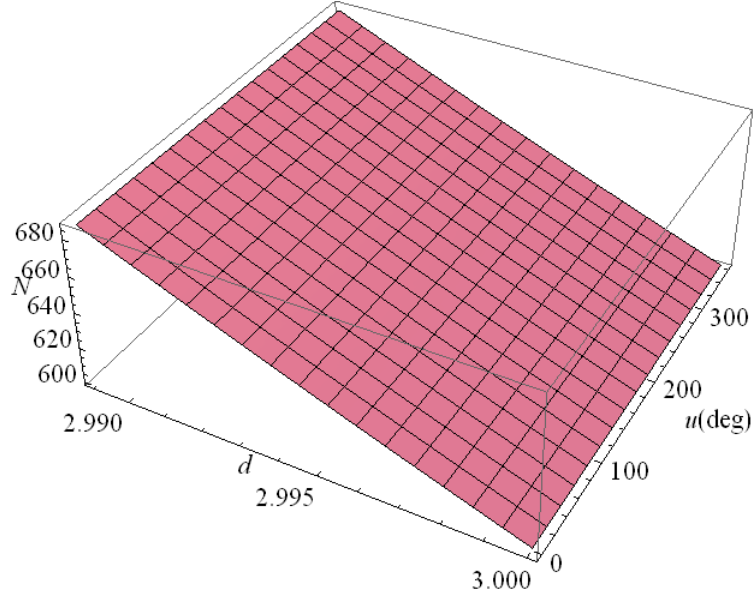

**Fig. 8** 3D plot of monomer number $N$ as a function of fractal dimension parameter $d$ and argument of latitude $u$ of a spacecraft in orbit for an experiment taking place in a spacecraft 300 km above the surface of the Earth in a circular polar orbit of inclination $i = 90°$ and eccentricity $e = 0$.

Using equation (13) we solve for the monomer minimum radius which for an Earth based experiment can be written as:

$$r_{min} = \left(\frac{N}{\xi}\right)^{1/d} + \Delta r_{cor_{Earth}} \tag{32}$$

$$\Delta r_{cor_{Earth}} = \frac{4\delta R_E^2 \left(f_E' \sin^2 \phi_E - 1\right)^2}{4GM_E - 4\gamma R_E^2 + 6GM_E J_{2_E} + 2Q'}, \tag{33}$$

where $\Delta r_{cor}$ is the correction to the minimum radius to the correction in gravitational acceleration.

$$Q' = 4\gamma f_E' R_E^2 - 9GM_E J_{2_E} \sin^2 \phi_E - 4 f_E'^2 R_E^2 \gamma \sin^4 \phi_E$$
$$+ 4\omega_E^2 R_E^3 \cos^2 \phi_E \left(-1 + f_E'^2 \sin^4 \phi_E \left(-3 + f' \sin^2 \phi_E\right)\right) + 3 f_E' R_E^3 \omega_E^2 \sin^2 2\phi_E, \tag{34}$$

On the surface of the Earth we obtain the following results tabulated in table 5 below:

**Table 5** Minimum radius correction for various geocentric latitudes on the Earth using $d$ =3.0 and $N$= 599.979.

| Experiment geocentric latitude $\phi_{Earth}$ | Minimum radius correction term $\Delta r_{cor}$ [pm] |
|:---:|:---:|
| 0 | 0.7899 |
| 30 | 1.5808 |
| 60 | -1.5819 |
| 90 | -0.7914 |

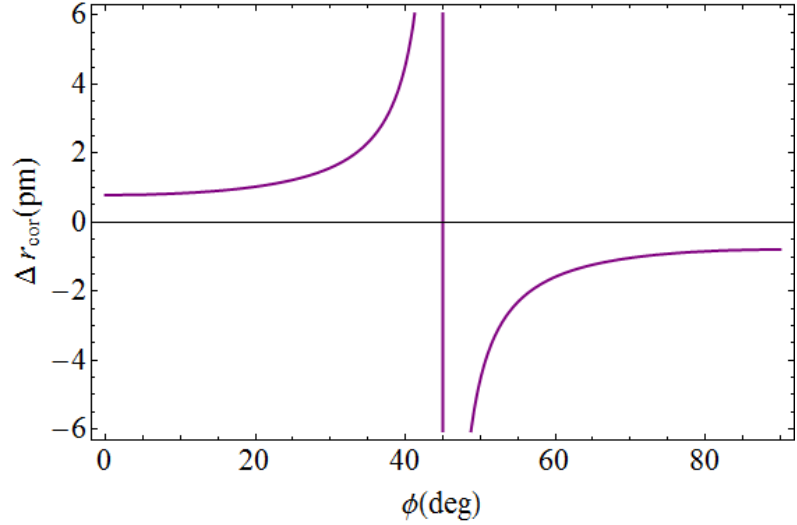

**Fig. 9** Plot of the closed packed monomers distance correction $\Delta r_{min}$ as a function of geocentric latitude $\phi$ using $d = 3.0$ and $N = 599.979$.

For an experiment above an orbiting spacecraft in a similar way we obtain:

$$r_{min} = \left(\frac{N}{\xi}\right)^{\frac{1}{d}} + \Delta r_{cor}, \qquad (35)$$

where $\Delta r_{cor}$ is the correction to the minimal radius due to gravitational acceleration correction at orbital point and it's given by:

$$\Delta r_{cor} = \frac{-\delta}{\gamma + H - Q''} = \frac{-\delta}{\gamma - \frac{GM_E\left(1 + e\cos f\right)^2}{a_s^2\left(1 - e^2\right)^2} + \frac{3GM_E R_E^2 J_2\left(1 + e\cos f\right)^4}{2a_s^4\left(1 - e^2\right)^4}\left(3\sin^2 i\sin^2 f - 1\right)} \qquad (36)$$

where $\Delta r_{cor}$ is the correction to the minimum radius to the correction in gravitational acceleration in orbit around the Earth. Our results are tabulated in table 5 below:

**Table 6** Minimum radius correction due to a corrected gravitational acceleration for an experiment taking place in a spacecraft in orbit around Earth at 300 km using $d = 3.0$ and $N = 599.979$, $e = 0.01$.

| Orbital inclination $i$ | Minimum radius correction term $\Delta r_{cor}$ [pm] |
|:---:|:---:|
| 0 | 0.0243 |
| 30 | 0.0240 |
| 60 | 0.0234 |
| 90 | 0.0232 |

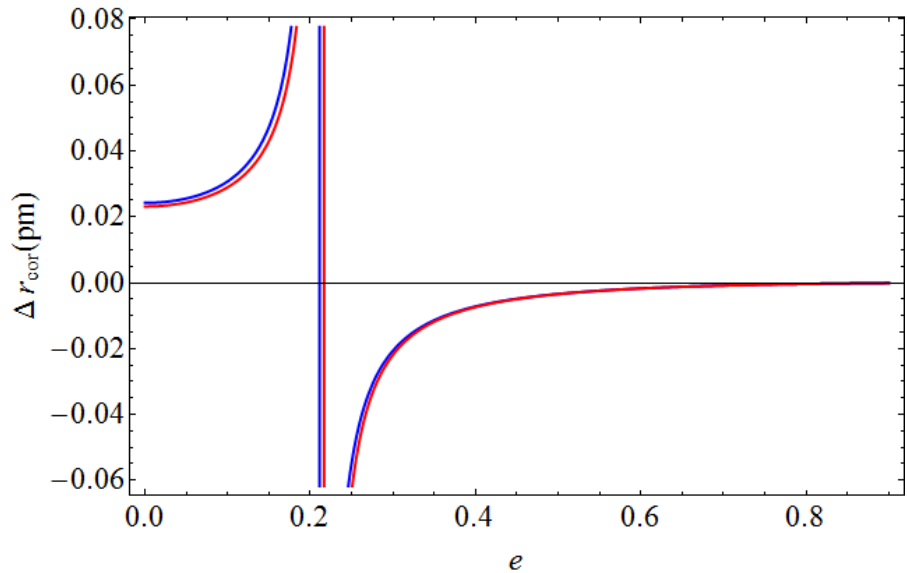

**Fig. 10** Minimum radius correction due to a corrected gravitational acceleration around
the orbit for an experiment taking place in a spacecraft in orbit around the Earth at 300 km
using $d = 3.0$ and $N = 599.979$, $e = 0.01$. The blue and red curves correspond to orbital inclinations
$i = 0°$ and $90°$ correspondingly.

With reference to table1, using $d = 3.0$ and constant gravity value $g \approx 9.81$ m/s² we find that the number of monomers $N$ is equal to 599.979. Calculating the number of monomers $N$ in the geocentric latitude range $0° \leq \phi_E \leq 90°$ we find a 0.001% difference between the equator and the pole, for which we can write that $N(\phi = 0°) = 0.9999 N(\phi = 90°)$. Similarly, the number of monomers $N$ when calculated for a smaller fitting parameter $\delta = 2.989$ and constant gravity $g$, results to 13% difference when compared to the value calculated under corrected gravity, where at the same time the 0.001% difference exist between equator and pole values under corrected gravity. In table 2 under Martian constant and corrected gravity we find there is no significant effect in the number of monomers $N$ which appears to be identical to that of the Earth. Furthermore, tables 4 and 5 give numerical results in orbit around Earth at the orbital altitude $h = 300$ km for various eccentricities and inclinations in the range $0 \leq e \leq 0.7$ and $0° \leq i \leq 90°$ respectively. Thus we find that there is no significant result, and that we conclude that the orbital elements of inclination and eccentricity as well as the $J_2$ harmonic do not significantly affect the number of monomers $N$. No significant effect exists between circular and elliptic orbits either. Finally, in tables 6 and 7 we have calculated the theoretical changes in minimum value $r_{min}$ or which the monomers are close-packed. From our analysis we see that corrected gravity on the surface of a planet or in orbit around it (in our case Earth) introduces a correction, which when calculated is minute in the order of picometers (i.e 1 pm = $10^{-12}$ m) On one hand the correction on the surface of the Earth depends on the $J_2$, harmonic, angular velocity of the Earth $\omega_E$, the fattening of the Earth $f'$, the radius and mass of the Earth $R_E$ and $M_E$ as well as the fitting parameters $\gamma$ and $\delta$. On the other surface on the Earth the contribution to the correction of $r_{min}$ becomes positive and negative depending on the geocentric latitude, increasing or decreasing thus minimum value $r_{min}$ of the closed packed monomers.

In figure 4 we plot the monomer number $N$ as function closed packed monomers distance $r_{min}$ and a fractal dimension parameter $d$ =3.0 for an experiment taking place at geocentric latitude $\phi = 90^o$ on the surface of the Earth. We see that the monomer number $N$ increases as $r_{min}$ increases according to the relation with $N \propto \xi\, r_{min}^d \propto \xi\, r_{min}^3$ being the leading term. For $d = 3$ no monomers form i.e. $N = 0$ when $r_{min}$ takes the following values below which are the roots of the following equation:

$$\xi r_{min}^3 + \frac{3\xi\delta}{g+\gamma} r_{min}^2 + \frac{3\xi\delta^2}{(g+\gamma)^2} r_{min} + \frac{\xi\delta^3}{(g+\gamma)^3} = 0 \,. \tag{37}$$

Solving Eq. (34) we obtain the following roots:

$$r_{1\,min} = r_{2\,min} = r_{2\,min} = -\frac{\delta}{(g+\lambda)}\,, \tag{38}$$

Similarly, for $d = 3$ but $N \neq 0$ solving the equation:

$$\xi\, r_{min}^3 + \frac{3\xi\delta}{g+\gamma} r_{min}^2 + \frac{3\xi\delta^2}{(g+\gamma)^2} r_{min} + \frac{\xi\delta^3}{(g+\gamma)^3} = N \,, \tag{39}$$

we obtain the following roots:

$$r_{min_1} = -\frac{\delta}{(g+\gamma)} - \left(\frac{N}{\xi}\right)^{1/3}\,, \tag{40}$$

$$r_{min_2} = -\frac{\delta}{(g+\gamma)} - \frac{(1-i\sqrt{3})}{2}\left(\frac{N}{\xi}\right)^{1/3}\,, \tag{41}$$

$$r_{min_2} = -\frac{\delta}{(g+\gamma)} - \frac{(1+i\sqrt{3})}{2}\left(\frac{N}{\xi}\right)^{1/3}\,. \tag{42}$$

Similarly, if $N$, $\gamma$ and $r_{min}$ are known the parameter $\delta$ for $N = 0$ and $N \neq 0$ are given by the equations:

$$\delta_1 = \delta_2 = \delta_3 = -r_{min}(g+\gamma)\,, \tag{43}$$

$$\delta_1 = -(g+\gamma)\left(r_{min} + \left(\frac{N}{\xi}\right)^{1/3}\right)\,, \tag{44}$$

$$\delta_2 = -(g+\gamma)\left(r_{min} + \frac{(1-i\sqrt{3})}{2}\left(\frac{N}{\xi}\right)^{1/3}\right)\,, \tag{45}$$

$$\delta_3 = -(g+\gamma)\left(r_{min} + \frac{(1+i\sqrt{3})}{2}\left(\frac{N}{\xi}\right)^{1/3}\right)\,. \tag{46}$$

In figure 5, we plot monomer number $N$ as a function of the spacecraft orbital inclination $i$ and eccentricity $e$ for two values of the fractal dimension parameter $d = 2.989$, $3.0$ respectively. Green plane corresponds to $d = 3.0$ and purple plane corresponds to $d = 2.989$. This is for an experiment taking place 300 km above the surface of the Earth. We find that a smaller dimensionality $d$ results to a higher number but constant monomer number $N$. Orbital inclination $i$ and eccentricity $e$ have no effect on the monomer number $N$. In figure 4 represents the monomer number $N$ as a function of the spacecraft orbital semimajor axis $a$ for a polar orbit of eccentricity $e = 0.1$ and inclination $i = 90°$ and for a fractal dimension parameter $d = 3.0$. We find that a change in the orbital semimajor axis change does not significantly affect the monomer number $N$, which remains constant instead. In figure 7, we plot monomers number $N$ as a function of spacecraft orbital semimajor axis $a$ and monomer minimum distance $r_{min}$, for an experiment taking place in a spacecraft 300 km above the surface of the Earth in an elliptical polar orbit of eccentricity $e = 0.1$ and inclination $i = 90°$ and for a fractal dimension parameter $d = 3.0$. We find that the monomer number $N$ increases as the monomer minimum distance $r_{min}$ increases with no important effect contribution visible as the orbital semimajor axis changes. In figure 8 we plot of monomer number $N$ as a function of fractal dimension parameter $d$ and argument of latitude $u$ of a spacecraft in a circular polar orbit for an experiment taking place in a spacecraft 300 km above the surface of the Earth with inclination $i = 90°$ and eccentricity $e = 0$. We find that there is no periodic effect in relation to the argument of latitude $u$, but rather that a smaller $d$ results to higher $N$. In figure 9 we plot the closed packed monomers distance corrections $\Delta r_{min}$ as a function of geocentric latitude $\phi$ using $d = 3.0$ and $N = 599.979$. We find that at the geocentric latitude $\phi = 45°$ the graph exhibits a vertical asymptote in the behaviour of the closed packed monomers distance correction $\Delta r_{min}$, being almost constant up to the geocentric latitude of $\phi \cong 20°$ contributing thus to a steady increase of closed packed monomers distance $r_{min}$ that reaches the approximate maximum value of 6 pm. Finally, at approximately $\phi = 55°$ the closed packed monomers distance correction $\Delta r_{min}$ becomes negative contributing thus to a steady reduction of closed packed monomers distance $r_{min}$. Finally, in fig. 10 we plot the minimum closed packed monomers the minimum closed packed monomers distance correction $\Delta r_{min}$ as a function of the orbital eccentricity $e$ for two different orbital inclinations namely $i = 0°$ and $90°$, i.e. equatorial and polar orbits correspondingly. In a similar way there exists a vertical asymptote approximately at the eccentricity values $e \cong 0.22 - 0.23$, where at approximately $e = 0.05$ the minimum closed packed monomers distance correction $\Delta r_{min}$ contributes positively to closed packed monomers distance $r_{min}$, decreasing progressively at $e \cong 0.25$ reducing thus $\Delta r_{min}$ and therefore $r_{min}$. Therefore we conclude that the geocentric latitude and eccentricity has a periodic increasing decreasing contribution on closed packed monomers distance correction $\Delta r_{min}$ which affects the closed packed monomers distance $r_{min}$, in the same way.

## 8. Conclusions

We have investigated the fractal structure on the surface of the Earth and also in orbit around it. In particular we have corrected the gravitational acceleration on the surface of the Earth's for oblateness harmonic coefficient $J_2$ and rotation angular velocity $\omega_E$ where in space just for the $J_2$ since the rotation of the Earth does not affect the spacecraft. In particular we have investigated the number of monomers $N$ on the Earth's surface as well as in orbit. Using the fitting parameter $d$ =3.0 we have found that the aggregate monomer number $N$ is not significantly affected and exhibits a minute 0.0001% difference between the geocentric and areocentric latitudes of $90^{\circ}$ and $0^{\circ}$. Finally for circular and elliptical orbits around Earth and Mars of various inclinations and eccentricities the aggregate monomer number it's not affected at all at the orbital altitude of 300 km. On the other hand our analysis shows that corrected gravity on the surface of a planet or in orbit around it (in our case Earth) introduces a correction, which when calculated is minute in the order of picometers (i.e 1 pm = $10^{-12}$ m). Moreover on the surface on the Earth the contribution to the correction of $r_{min}$ becomes positive and negative depending on the geocentric latitude, increasing or decreasing thus minimum value $r_{min}$ of the closed packed monomers total distance.

**Acknowledgments** The authors would like to thank two anonymous reviewers, who with their encouraging and constructive comments helped us improve the manuscript.